\documentclass[aps,prd,twocolumn,superscriptaddress,showpacs,a4,twoside]{revtex4}
\usepackage{graphicx,axodraw,bm,amssymb,amsmath,subfigure}

\begin{document}

\newcommand{\beq}{\begin{equation}}
\newcommand{\eeq}{\end{equation}}

\title{Interference between $\bm{f_0(980)}$ and $\bm{\rho(770)^0}$ resonances in $\bm{B \to \pi^+\pi^- K}$ decays}

\author{B.~El-Bennich}
\affiliation{Laboratoire de Physique Nucl\'eaire et de Hautes \'Energies (IN2P3--CNRS--Universit\'es Paris 6 et 7), Groupe Th\'eorie, 
                  Universit\'e Pierre et Marie Curie, 4 place Jussieu, 75252 Paris, France}
\author{A.~Furman}
\affiliation{ul. Bronowicka 85/26, 30-091 Krak\'ow, Poland}
\author{R.~Kami\'nski} 
\author{L.~Le\'sniak}
\affiliation{Division of Theoretical Physics, The Henryk Niewodnicza\'nski Institute of Nuclear Physics,
                  Polish Academy of Sciences, 31-342 Krak\'ow, Poland}
\author{B.~Loiseau}
\affiliation{Laboratoire de Physique Nucl\'eaire et de Hautes \'Energies (IN2P3--CNRS--Universit\'es Paris 6 et 7), Groupe Th\'eorie,
                  Universit\'e Pierre et Marie Curie, 4 place Jussieu, 75252 Paris, France}
\date{\today }

\begin{abstract}
We study the contribution of the strong interactions between the two pions in $S$- and $P$-waves to the weak $B \to \pi\pi K$ decay amplitudes. 
The interference between these two waves is analyzed in the $\pi\pi$ effective mass range of the $\rho(770)^0$ and $f_0(980)$ resonances.  
We use a unitary $\pi\pi$ and $\bar KK$ coupled-channel model to describe the $S$-wave interactions and  a Breit-Wigner function for the 
$P$-wave amplitude. The weak $B$-decay amplitudes, obtained from QCD factorization, are supplemented with charming penguin contributions 
in both waves. The four complex parameters of these long distance terms are determined by fitting the model to the BaBar and Belle data 
on $B^{\pm,0} \to \pi^+\pi^- K^{\pm,0}$ branching fractions, $CP$ asymmetries, $\pi\pi$ effective mass and helicity-angle distributions. 
This set of data, and in particular the large direct $CP$-asymmetry for $B^\pm \to \rho(770)^0 K^\pm$ decays, is well reproduced.
The interplay of charming penguin amplitudes and the interference of  $S$- and $P$-waves describes rather successfully the experimental 
$\mathcal{S}$ and $\mathcal{A}$ values of the $CP$-violating asymmetry for both $B^0 \to f_0(980) K_S^0$ and $B^0 \to \rho(770)^0 K_S^0$ decays.

\end{abstract}

\maketitle

\section{Introduction \label{intro}}

Hadronic final-state interactions in three-body charmless $B$ meson decays supply good opportunities for $CP$ violation searches. In order to 
interpret in the most reliable way decay observables, it is necessary to take into account strong interaction effects between the produced meson 
pairs. These effects, at relative energies below and about 1~GeV, are clearly visible in the Dalitz plots of three-body weak decays such as 
$B\to \pi\pi K$ and $B\to \bar KKK$ \cite{Garmash:2005rv,Garmash2005,Abe0509047,AubertPRD72,AubertPRD73,Aubert0408}.  In particular, one observes a 
distinct surplus of events at these relatively small effective masses distributed near the edge of the Dalitz plot. Prominent maxima are seen in the 
effective $\pi\pi$ mass distribution, notably the $f_0(980)$ and $\rho(770)^0$ resonances. Several resonances at higher energies are visible too, 
though their identification is less straightforward. The experimental effective mass distributions of three-body $B$-decays are extracted from Dalitz 
plots. One cannot describe these data without including final state interactions. The values of branching ratios obtained in experimental analyzes 
are model dependent. In phenomenological studies of $B$ meson decays, the isobar model is employed, which has a large number of arbitrary phases 
and relative intensity parameters. It is important to use a well-constrained model to describe the final state interactions, where the constraints 
can come from theory and analyzes of other processes in which the same final states are produced. One of our
aims is to reduce the number of free 
parameters by using a model which describes many resonances in a unitarized way.

In Ref.~\cite{fkll}, particular attention was paid to the $B \to f_0(980)K$ channels which have relatively large branching ratios for charmless 
$B$-decays. The amplitudes considered in \cite{fkll}, based on the QCD factorization approach, do not include hard-spectator and annihilation terms 
which contain phenomenological parameters. It was found that these amplitudes do not reproduce the $B^\pm \to f_0(980)K^\pm$ and 
$B^0 \to f_0(980)K^0$ branching ratios. Therefore, additional terms were included in the decay amplitudes in the form of long-distance contributions 
stemming from so-called charming penguins~\cite{Ciuchini:1997hb}. 
At the hadronic level, these contributions could be associated with intermediate $D_s^{(*)}D^{(*)}$ states and be understood as being part of the 
intricate final-state interactions in $B$ decays. One can expect a sizable contribution from such processes as $B\to D_s^*D^{(*)}$ branching fractions 
are large. The addition of charming penguins and of $\pi\pi$ and $\bar K K$ coupled-channel final state interactions in Ref.~\cite{fkll} allowed 
to obtain a good agreement with the measured $B \to f_0(980) K$ branching ratios and to reproduce the effective $\pi\pi$ mass distribution of the 
BaBar and Belle data. 

Long distance contributions, similar to the charming penguin effects, have been considered by Cheng, Chua and Soni~\cite{Cheng:2004ru}.
Cheng, Chua and Yang have recently studied the nature of the scalar mesons $f_0(980)$ and $a_0(980)$ in the decays $B\to f_0(980)K$, 
$B\to a_0(980)\pi$ and $B\to a_0(980)K$ using QCD factorization for the weak decay amplitudes~\cite{Cheng:2005nb}. Their calculations 
are for a $\bar qq$ state of the $f_0(980)$, but implications of a $q^2 \bar q^2$ picture of the $f_0(980)$ are also discussed. In their
work~\cite{Cheng:2005nb}, however, they neither consider hadronic long-distance contributions nor pionic and kaonic final-state interactions.

In the literature, one finds different results on the relative size of the weak annihilation or hard-spectator contributions to the $B$-decay amplitude
into two mesons~\cite{Arnesen:2006vb,Bauer04,BBNSnotes}. A recent study in soft-collinear effective theory~\cite{Arnesen:2006vb}  concludes that 
the annihilation and chirally enhanced annihilation contributions in charmless $B\to M_1 M_2$ decays are real to leading order in 
$\Lambda_\mathrm{QCD}/m_b$ (here $M_1$ and $M_2$ are non-isosinglet mesons, $\Lambda_\mathrm{QCD}$ is the QCD scale and $m_b$ the 
$b$-quark mass).  These contributions constitute a small fraction of the experimentally determined total penguin amplitudes in the case 
$\bar B^0 \to K^-\pi^+$ or $B^- \to K^- \bar K^0$ decays and in the present work they are not considered.  Even upon inclusion of the relatively small 
annihilation and hard-spectator contributions, the current QCD factorization results underestimate the average branching ratio for $B\to \rho K$ by a 
factor of two and larger~\cite{Leitner:2004ij,bene03} unless certain model parameters are strongly modified (see scenarios S2 to S4 in Ref.~\cite{bene03}). 
To be more quantitative, let us cite the branching ratio of the $B^- \to K^- \rho^0$ decay calculated by Leitner, Guo and Thomas~\cite{Leitner:2004ij}. 
Including fully these two contributions, it is equal to $1.54\times 10^{-6}$. This should be compared to the experimental branching ratio for this decay 
which lies between $3.9\times 10^{-6}$ and $5.1\times 10^{-6}$ with a typical error of $0.5\times 10^{-6}$~\cite{Garmash:2005rv,AubertPRD72}.

In this paper, we supplement the $B\to \rho(770)^0 K$, $\rho(770)^0 \to \pi^+\pi^-$  decay amplitudes to the $B\to f_0(980) K$, $f_0(980)\to\pi^+\pi^-$ 
amplitudes derived in Ref.~\cite{fkll}. The experimental branching ratio for the decay $B\to \rho(770)^0 K$ is of the same order of magnitude 
as for the $B\to f_0(980)K$ channel. A characteristic feature of the $B^\pm\to \rho(770)^0 K^\pm$ decay, however, is the very large 
value of the direct $CP$ violating asymmetry of about $0.3$ measured by Belle~\cite{Garmash:2005rv} and BaBar~\cite{AubertPRD72}. 
The addition of the $B\to \rho(770)^0 K$ decay amplitudes enables us to study possible interferences between the $S$- and $P$-waves in 
$B\to \pi\pi K$ decays. These effects can be observed in particular in the time dependent $CP$ asymmetries of neutral $B$ decays 
into $\rho(770)^0K_S^0$ and $f_0(980)K_S^0$ channels. In $b\to s$ transitions, \textit {new physics} contributions to these asymmetries have 
been put forward~\cite{Buchalla:2005us,Beneke:2005pu}. Nonetheless, one should also take into account the influence of meson-meson final-state 
interactions on these observables, among others on the deviation of the asymmetry parameter $\mathcal{S}$ from the Standard 
Model expectation value $\sin 2\beta$.

In Section~\ref{sec2}, after a concise summary of the $S$-wave contributions $B\to f_0(980) K$ of Ref.~\cite{fkll}, we derive the $B\to \rho(770)^0 K$ 
decay amplitudes from the QCD factorization approach of Ref.~\cite{bene03}. We introduce a Breit-Wigner vertex function $\Gamma_{\rho\pi\pi}(m_{\pi\pi})$ 
in the $B\to \rho(770)^0\,K$ decay amplitude to account for the $\pi\pi$  final-state interactions in the $P$-wave. We complement this amplitude with 
the $S$-wave contribution $B\to f_0(980) K$ obtained in Ref.~\cite{fkll}. In Section~\ref{sec3}, we give our model expressions for the observables measured 
by the Belle and BaBar Collaborations. We briefly discuss our fitting method in Section~\ref{sec4}. In Section~\ref{sec5} our results are compared 
to the experimental $\pi\pi$ mass and helicity-angle distributions and we give our branching ratio values for $B\to \rho(770)^0 K$ and $B\to f_0(980)^0 K$ 
decays. We analyze the $S$- and $P$-wave interference effects as well as the influence of charming penguins on the $CP$ violating parameters 
$\mathcal{S}(m_{\pi\pi})$ and $\mathcal{A}(m_{\pi\pi})$ of the time-dependent asymmetry in neutral $B^0$ decays. Furthermore, we calculate the 
asymmetry parameters $\mathcal{S}$ and $\mathcal{A}$ and compare our results with experimental data. In Section~\ref{sec6}, we summarize and propose 
further improvements of this work. Finally, for completeness, we give, in the Appendix the expressions of the $B^-\to f_0(980)K^-$ and 
$B^-\to(\pi^+\pi^-)_{S-wave}K^-$ decay amplitudes. This will also allow us to make a comparison with other approaches like that of Ref.~\cite{Cheng:2005nb}.

\section{Amplitudes for the $\bm{B\to \pi\pi K}$ decays \label{sec2}}

\subsection{A brief recall of the $\bm{B\to f_0(980) K}$ amplitudes} 

In Ref.~\cite{fkll}, the $B$ decays into $\pi\pi K$ and $\bar KKK$ were studied for final-state $(\pi\pi)_S$ and $(\bar KK)_S$ pairs interacting 
in an isospin zero $S$-wave from $\pi\pi$ threshold to about 1.2~GeV. The QCD factorization approach is the framework used to describe weak 
decays of $B$ mesons into a quasi two-body state $f_0(980)K$. 
The effective coefficients $a_i(\mu) (i=1,...,6)$ at the renormalization scale 
$ \mu= 2.1$ GeV were taken from Ref.~\cite{deGroot:2003ms}. 
Corrections arising from annihilation topologies and hard gluon scattering 
with the spectator quark were not included. 

The two-pion and two-kaon rescattering effects are described by the $\pi\pi$ and $\bar KK$ unitary coupled-channel model of 
Ref.~\cite{Kaminski:1997gc}, which allows to describe the lowest scalar-isoscalar resonances $f_0(600)$ and $f_0(980)$ with
a single matrix of amplitudes. The meson-meson amplitudes are incorporated into four scalar form factors responsible for
the production of the $\pi\pi$ and $\bar KK$ pairs in an isospin zero $S$-wave from the current $\bar uu, \bar dd$ and $\bar ss$ 
antiquark-quark states. These scalar form factors are constrained by the chiral dynamics of low-energy meson-meson interactions
\cite{Meissner:2000bc} and provide the link between the weak decay amplitudes~\cite{deGroot:2003ms} and the hadronic rescattering 
model~\cite{Kaminski:1997gc}.

As mentioned in the introduction, charming penguin amplitudes were included. In a nutshell, these amplitudes are low-momentum $\bar cc$ 
loop contributions which are formally suppressed by powers of $\Lambda_{\mathrm{QCD}}/m_b$ though numerically enhanced by large Wilson 
coefficients~\cite{Ciuchini:1997hb}. No explicit calculations being available so far, one interprets these amplitudes as long distance contributions 
parameterized by two complex parameters.

\subsection{The $\bm{B\to \rho K, \rho \to (\pi\pi)_P}$ decay amplitudes}

\begin{center}
\begin{figure}[h!]
\subfigure{\includegraphics*[width=0.32\textwidth]{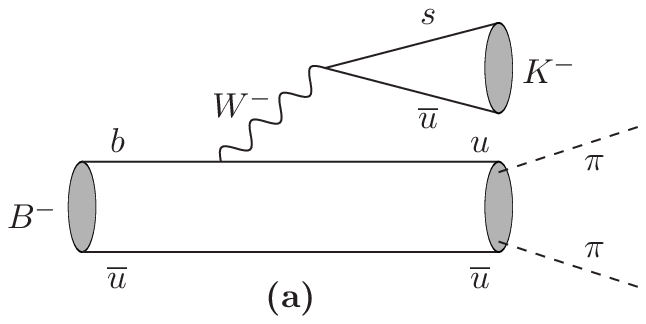}}\\
\subfigure{\includegraphics*[width=0.32\textwidth]{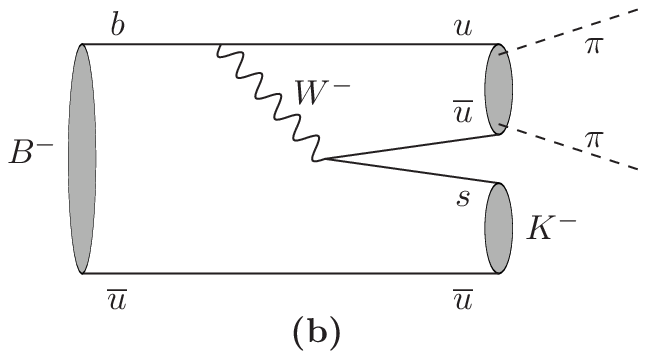}}\\
\subfigure{\includegraphics*[width=0.32\textwidth]{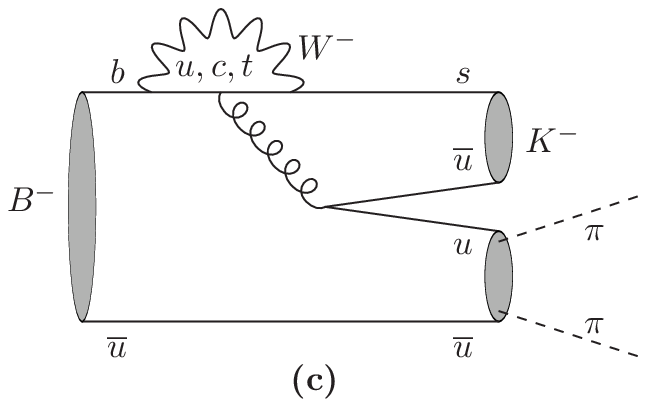}}\\
\subfigure{\includegraphics*[width=0.32\textwidth]{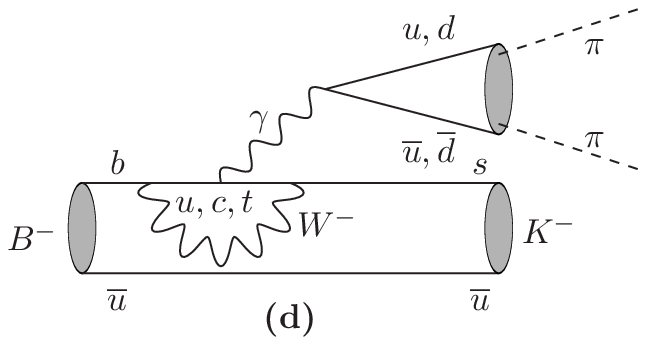}}\\
\subfigure{\includegraphics*[width=0.32\textwidth]{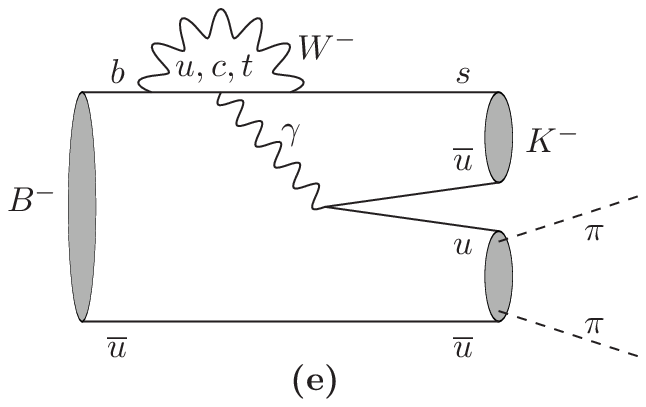}}
\caption{Quark line diagrams for the $B^-$ decay: {\bf (a)} and {\bf (b)} are tree diagrams, {\bf (c)} is the penguin diagram and {\bf (d)} and {\bf (e)} 
are electroweak penguin diagrams. The spring-like lines represent gluon exchange, wavy lines photon or $W^-$ exchange and the dashed ones the 
$\pi\pi$ isospin 1 $P$-wave pair.}
\label{fig1}
\end{figure}
\end{center}

The derivation of the $B\to \rho(770)^0\,K$ weak decay amplitude follows the QCD factorization approach of Beneke and Neubert~\cite{bene03}.
By $(\pi\pi)_P$ we denote isovector $\pi\pi$ states in a $P$-wave and in the following we will write interchangeably $\rho(770)^0$ or 
$(\pi\pi)_P$ depending on the context. The possible quark line diagrams for negative $B$-meson decays are shown in Fig.~\ref{fig1}, 
where the $\pi^+\pi^-$ final state interaction is graphically schematized by dashed lines appended to either a $\bar uu$ or $\bar dd$ 
state in the diagrams. 

The $B^{-} \to (\pi^+\pi^-)_P\,K^{-}$ decay amplitude can be written as
\beq
\langle(\pi^+\pi^-)_P\,K^-\vert H \vert B^-\rangle =
p_K \cdot \left(p_{\pi^-}-p_{\pi^+}\right) A^-\,\Gamma_{\rho\pi\pi}(m_{\pi\pi}),
\label{full.amplitude}
\eeq
where $p_K$ is four-momentum of $K^-$ and $p_{\pi^-}$ and $p_{\pi^+}$ are the four-momenta of the negative and positive pions, respectively. 
In the  $\pi^+\pi^-$ rest frame, the decay amplitude reduces to
\begin{eqnarray} 
\lefteqn{\langle(\pi^+\pi^-)_P\, K^- \vert H \vert B^- \rangle } \hspace*{1.2cm} \nonumber \\ 
 & & = 2 A^-\,\Gamma_{\rho\pi\pi}(m_{\pi\pi})\,|\mathbf{p}_{\pi}||\mathbf{p}_K| \cos \theta,
\label{fullbis.amplitude}
\end{eqnarray}
where $|\mathbf{p}_{\pi}|$ and $|\mathbf{p}_K|= \sqrt{E_K^2(m_{\pi\pi})\,-\,M_K^2} $ are the moduli of the pion and kaon momenta
with $E_K(m_{\pi\pi}) = \frac{1}{2}(M_B^2-m_{\pi\pi}^2-M_K^2)/m_{\pi\pi}$, $M_B$, $m_{\pi\pi}$ and $M_K$ being the $B$-meson, effective 
$\pi\pi$ and kaon masses, respectively. Moreover, $\theta$ is the helicity angle between the direction of flight of the $\pi^-$ in the
$\pi^+\pi^-$ rest frame and the direction of the $\pi^+\pi^-$ system in the $B$ rest frame. 

The function  $A^-$ is given by
\begin{eqnarray}
A^- &= &G_F\, m_{\rho} [\, f_K\,A_0^{B\to \rho}(M_K^2)\,( U^- - C^P ) \nonumber \\
 & + & f_{\rho}\,F_1^{B\to K}(m_{\rho}^2)\,W^- ],
\label{simplification.beneke}
\end{eqnarray}
where $G_F$ is the Fermi coupling constant, $f_K = 159.8$~MeV is the kaon decay constant and the $\rho$ decay constant is given by $f_\rho = 209$~MeV. 
For the  transition form factors $B\to \rho$ and $B\to K$, we employ $A_0^{B\to \rho}(M_K^2) = 0.37$ and $F_1^{B\to K}(m_{\rho}^2) = 0.34$, 
respectively. The above values for decay constants and transition form factors are taken from Table~I of Ref.~\cite{bene03}. The functions $U^-$ 
and $W^-$ are defined as
\begin{eqnarray}
\label{eq:U}
U^- &= &\lambda_u [\, a_1+a_4^{u} -a_4^c-r  \left ( a_6^u-a_6^c + a_8^u-a_8^c \right )  \nonumber \\
  & + & a_{10}^u-a_{10}^c \,] +\lambda_t  [\,-a_4^c+r (a_6^c+a_8^c)-a_{10}^c  \, ] 
\end{eqnarray}
and
\beq
  W^-  =  \lambda_u\,a_2 - \lambda_t  \mbox{$\frac{3}{2}$} (a_7+a_9).
\label{eq:W}
\eeq
The chiral factor is given by $r=2 M_K^2/[(m_b+m_u)(m_s+m_u)]$, with $m_b=4.46$~GeV and where  $m_u=5$~MeV and $m_s=110$~MeV are the  $u$- and 
$s$-quark masses, respectively. The coefficients $a_i(\mu =m_b/2)$ are taken to be the $a_{i,I}(\pi K)$ of Table~III in Ref.~\cite{bbns01}. 
In Eq.~(\ref{eq:U}), we did not include the annihilation contribution considered in Ref.~\cite{bene03}. The charming penguin contribution 
$C^P$ in Eq.~(\ref{simplification.beneke}) is parameterized as
\beq
C^P =  \lambda_u P_u + \lambda_t P_t ,
\label{def:C}
\eeq
with $P_u$ and $P_t $ being complex parameters \cite{Ciuchini:1997hb}. In Eqs.~(\ref{eq:U}), (\ref{eq:W}) and (\ref{def:C}),
we have used the unitarity condition $\lambda_c + \lambda_u + \lambda_t=0$, where $\lambda_u=V_{ub}V_{us}^*$, $\lambda_t=V_{tb}V_{ts}^*$ 
and $\lambda_c=V_{cb}V_{cs}^*$ are products of the  Cabibbo--Kobayashi--Maskawa (CKM) matrix elements, to express the amplitudes 
in terms of $\lambda_u$ and $\lambda_t$. The parameters of the CKM matrix are taken from  Ref.~\cite{Charles}.

The corresponding $A^+$ function of the $B^+ \to (\pi^-\pi^+)_P\,K^+$ decay amplitude is obtained from replacing 
the $\lambda_u$ and $\lambda_t$ values by their complex conjugates $\lambda_u^*$ and $\lambda_t^*$ in Eqs.~(\ref{eq:U}) 
to (\ref{def:C}) and changing the overall sign in Eq.~\eqref{fullbis.amplitude},
\begin{eqnarray}
\lefteqn{\langle(\pi^-\pi^+)_P\, K^+ \vert H \vert B^+ \rangle =} \hspace*{1.2cm} \nonumber \\ 
 & & = 2 A^+\,\Gamma_{\rho\pi\pi}(m_{\pi\pi})\,|\mathbf{p}_{\pi}||\mathbf{p}_K| \cos \theta,
\label{B+amplitude}
\end{eqnarray}
where we define
\beq
 A^+  = - A^-(\lambda_u^*,\lambda_t^*).
\label{A+}
\eeq
The minus sign in Eq.~\eqref{A+} follows from the $CP$ symmetry of the final state. 

Coming to neutral $B$ decays, the $\bar B^{0}\to \rho(770)^0 \bar K^{0}$, $\rho^0 \to \left(\pi^+\pi^- \right)_P$ decay amplitude reads
\begin{eqnarray}
\lefteqn{\langle(\pi^+\pi^-)_P\,\bar K^0\vert H\vert \bar B^0\rangle  } \hspace*{1.3cm}\nonumber  \\
& = & 2 \bar A^0\,\Gamma_{\rho\pi\pi}(m_{\pi\pi})\,|\mathbf{p}_{\pi}|
|\mathbf{p}_K| \cos \theta,
\label{full0.amplitude}
\end{eqnarray}
with
\begin{eqnarray}
 \bar A^0 &=& G_F\, m_{\rho} \left [ f_K\,A_0^{B\to \rho}(M_K^2)\,(\bar U^0+C^P) \right. \nonumber \\
  &+ &  \left. f_{\rho}\,F_1^{B\to K}(m_{\rho}^2)\, \bar W^0 \right ].
\label{simplification.beneke0}
\end{eqnarray}
The different sign in front of the charming penguin term $C^P$ is due to the $\bar dd$ quark content of the $\rho(770)^0$ in 
neutral $B$ decays in contrast to Eq.~\eqref{simplification.beneke}, where in charged $B$ decays the $\bar uu$ configuration 
 comes into play. The function $\bar U^0$ is given by
\begin{eqnarray}
\label{eq:U0}
\lefteqn{\hspace*{-6mm} \bar U^0 = \lambda_u \{ -a_4^{u} + a_4^c +r \left  [ a_6^u - a_6^c -  (a_8^u-a_8^c)/2 \right ] } \nonumber \\ 
 & & \hspace*{-5mm} +  ( a_{10}^u - a_{10}^c )/2 \} + \lambda_t [a_4^c-\! r (a_6^c-a_8^c/2)-a_{10}^c/2 ]
\end{eqnarray}
and $\bar W^0=W^-$. Here again we have left out  annihilation contributions as considered in Ref.~\cite{bene03}. The replacement 
of $\lambda_u$ and $\lambda_t$ by their complex conjugate values in Eqs.~\eqref{eq:W} and \eqref{eq:U0} leads to the 
corresponding function $A^0$  for the $B^0 \to (\pi^- \pi^+)_P\, K^0$  decays which reads
\begin{eqnarray}
 \lefteqn{\langle(\pi^-\pi^+)_P\, K^0\vert H\vert  B^0\rangle  } \hspace*{1.3cm}\nonumber  \\ 
 & = & 2 A^0\,\Gamma_{\rho\pi\pi}(m_{\pi\pi})\,|\mathbf{p}_{\pi}| |\mathbf{p}_K| \cos \theta,
\label{full0bar.amplitude}
\end{eqnarray}
where one has again to account for an overall sign change in the amplitude
\beq
  A^0  =  - \bar A^0(\lambda_u^*,\lambda_t^*).
\label{A0}
\eeq

The $\rho \to \pi\pi$ vertex function is chosen to be of Breit-Wigner form,
\beq
\Gamma_{\rho\pi\pi}(m_{\pi\pi}) = \frac{g_{\rho\pi\pi}}{m_{\pi\pi}^2-m_{\rho}^2+
i\,\Gamma_{\rho}\,m_{\rho}},
\label{Gamma.rho}
\eeq
where $m_{\rho}=775.8$~MeV, the decay width is $\Gamma_{\rho}=150.3$~MeV and the coupling constant 
is taken to be $g_{\rho\pi\pi}^2/4\pi = 3\, m_{\rho}^2\, \Gamma_{\rho}/ 2 p_\pi^3$.

\subsection{Complete $B\to \pi\pi \,K$ decay amplitudes\label{sec2c}}

We add the $S$-wave amplitude $a_S$ of Ref.~\cite{fkll} for the $B\to (\pi\pi)_S\,K$ decays to obtain the complete decay 
amplitude as
\beq
\mathcal{M} =  a_S + a_P|\mathbf{p}_{\pi}| |\mathbf{p}_K|\cos \theta,
\label{largeM}
\eeq
where $a_S$ is given by either Eq.~(1) or Eq.~(7) of Ref.~\cite{fkll} and $a_P$ can be read either from Eq.~\eqref{fullbis.amplitude} or 
Eq.~\eqref{full0.amplitude} for $B^-$ decays  or for $\bar B^0$ decays, respectively. The expression of the $B^-\to(\pi^+\pi^-)_SK^-$ amplitude 
can also be found in the Appendix \ref{appendix} (Eq.~(A12)). It follows from Ref.~\cite{fkll} and from Eqs.~\eqref{A+} and \eqref{A0} 
that the $CP$-conjugate $S$- and $P$-wave amplitudes $\bar a_S$ and $\bar a_P$ are related to the amplitudes $a_S$ and $a_P$ by
\beq
  \bar a_S = a_S(\lambda_u^*,\lambda_t^*), \quad \bar a_P = - a_P(\lambda_u^*,\lambda_t^*).
\eeq 
Thus, the conjugate of $\mathcal{M}$ is written as
\beq
  \mathcal{\bar M} = \bar a_S + \bar a_P|\mathbf{p}_{\pi}| |\mathbf{p}_K|\cos \theta.
\label{largeMbar}
\eeq
In the amplitude $a_S$ of the present work, the charming penguin contribution $C(m)$ of  Ref.~\cite{fkll} is replaced by  
\beq
  C^S(m) = - (M_B^2-m^2) f_K F_0^{B \to (\pi \pi)_S}(M_K^2)
(\lambda_u S_u + \lambda_t S_t).   
\label{newCharmS} 
\eeq
In Eq.~(\ref{newCharmS}), $S_u$ and $S_t$ are two complex parameters and $F_0^{B \to (\pi \pi)_S}(M_K^2)=0.46$~\cite{fkll} is the $B$-transition 
form factor to a pair of pions in an $S$-state. Furthermore, the constant $\chi$, characteristic of the $f_0(980) \to \pi\pi$ decay, is fixed to 
29.8 GeV$^{-1}$ in agreement with the estimation made in Ref.~\cite{fkll}. 
Here, for the $S$-wave, we also use 
the $a_{i,I}(\mu=m_b/2)$ from Table III of Ref.~\cite{bbns01} as effective coefficients $a_i$.

We can substitute the coefficients $a_i(PP)$ of the $B$-decay into two pseudoscalar mesons $PP$ in place of the coefficients $a_{i,I}(PS)$ corresponding 
to the final pseudoscalar-scalar $PS$ state for the following reasons. As can be seen from Eq.~(35) of Ref.~\cite{bene03}, the vertex corrections, 
which depend on the scalar light-cone distribution amplitudes, affect only the $a_{6,I}(PS)$ coefficients of the amplitude for which the emitted 
meson without the quark spectator is a scalar. In our $S$-wave amplitude it is only found in the $a_6$ of Eq.~(A5). However, $a_{6,I}(PS)$ is 
practically equal to $a_{6,I}(PP)$ (see Ref.~[10] and Ref.~[11] cited in Ref.~\cite{fkll}). Any further corrections to the $a_i$ are due to 
hard-gluon scattering contributions ($a_{i,II}(\mu)$ of Ref.~\cite{bbns01}) as well as annihilation terms; both processes are, however, not 
included here.

\section{Definitions of observables\label{sec3}}

The $B\to \pi\pi K$ amplitude $\mathcal{M}$ in Eq.~\eqref{largeM} depends on the effective mass $m_{\pi\pi}$ and $\cos\theta$ 
which is equivalent to the other effective mass variable $m_{\pi K}$ on the Dalitz plot. The double differential $B\to \pi\pi K$ decay 
distribution then reads
\beq
\frac{d^2 \Gamma}{d\cos \theta\, dm_{\pi\pi}} = \frac{m_{\pi\pi}|\mathbf{p}_\pi ||\mathbf{p}_K|}{8\,(2\pi)^3\,M_B^3} |\mathcal{M}|^2,
\label{doubledecay.rate}
\eeq
Integrating over $\cos\theta$ one obtains the differential $B\to \pi\pi K$ decay distribution
\beq\hspace*{-2mm}
\frac{d\Gamma}{dm_{\pi\pi}} = \frac{m_{\pi\pi}|\mathbf{p}_\pi | |\mathbf{p}_K|}{4\,(2\pi)^3\,M_B^3}
               \left ( |a_S|^2  + \frac{1}{3} |\mathbf{p}_\pi |^2 |\mathbf{p}_K |^2 |a_P|^2 \right )
\label{mpipidecay.rate}
\eeq
and the differential branching ratio is given by 
\beq
  \frac{d\mathcal{B}}{dm_{\pi\pi}} = \frac{1}{\Gamma_B} \frac{d\Gamma}{dm_{\pi\pi}} ,
\label{diffbranchratio}  
\eeq
where $\Gamma_B$ is the appropriate total width for $B^+$ or $B^0$.

The integration on the Dalitz plot over the kinematically allowed $m_{\pi\pi}$ range yields the helicity angle $B\to \pi\pi K$ 
distribution
\beq
\frac{d\Gamma}{d\cos \theta}=A + B\cos\theta + C\cos^2\theta.
\label{dGamdcos}
\eeq
The functions $A$, $B$ and $C$ are defined as
\begin{eqnarray}
A &=& \int_{m_{\mathrm{min}}}^{m_{\mathrm{max}}}\frac{m_{\pi\pi}|\mathbf{p}_{\pi}||\mathbf{p}_K |}{8\,(2\pi)^3\,M_B^3}|a_S|^2 \ dm_{\pi\pi},
\label{largeA} \\
B &=& 2 \int_{m_{\mathrm{min}}}^{m_{\mathrm{max}}}\frac{m_{\pi\pi}|\mathbf{p}_{\pi}|^2 |\mathbf{p}_K|^2}{8\,(2\pi)^3\,M_B^3}
\,\mathrm{Re}\, (a_S a_P^*) \ dm_{\pi\pi},
\label{largeB}  \hspace*{4mm}\\
C &=& \int_{m_{\mathrm{min}}}^{m_{\mathrm{max}}}\frac{m_{\pi\pi}|\mathbf{p}_{\pi}|^3 |\mathbf{p}_K|^3}{8\,(2\pi)^3\,M_B^3}
|a_P|^2 \ dm_{\pi\pi}.
\label{largeC}
\end{eqnarray}

The $CP$ violating asymmetry for charged $B$ decays is defined as  
\beq
\mathcal{A}_{CP} = \frac{\dfrac{d\Gamma (B^-\to \pi^+\pi^-K^-)}{dm_{\pi\pi}} - \dfrac{d\Gamma (B^+\to \pi^+\pi^-K^+)}{dm_{\pi\pi}}}
                                    {\dfrac{d\Gamma (B^-\to \pi^+\pi^-K^-)}{dm_{\pi\pi}} + \dfrac{d\Gamma (B^+\to \pi^+\pi^-K^+)}{dm_{\pi\pi}}}.
\label{acp}
\eeq
Furthermore, in neutral $B$ meson decays into final $CP$ eigenstates $f$ the time-dependent asymmetries are given by
\begin{eqnarray}
\mathcal{A}_{CP} (t) & = & \frac{\Gamma(\bar B^0(t)\to f) - \Gamma ( B^0(t)\to f)}{ \Gamma(\bar B^0(t)\to f) +\Gamma( B^0(t)\to f)} \nonumber \\
                                     & = & \mathcal{S} \sin(\Delta m t) + \mathcal{A}\cos(\Delta m t).
\label{acpt}
\end{eqnarray}
The mass difference between the two neutral $B$ eigenstates is denoted by $\Delta m$, $\mathcal{S}$ is a measure for the mixing induced 
$CP$ asymmetry, and $\mathcal{A}$ is the direct $CP$ violating asymmetry. The two parameters of the time dependent asymmetry, $\mathcal{S}$ 
and $\mathcal{A}$, are
\begin{subequations}
\begin{eqnarray}
  \mathcal{S} & = &\frac{2\,\mathrm{Im}\, \lambda_f}{1+|\lambda_f |^2} = (1-\mathcal{A} )\,\mathrm{Im}\, \lambda_f, \label{Sf}
\\ 
  \mathcal{A} & = &- \frac{1-|\lambda_f |^2}{1+|\lambda_f |^2},
\label{Af}
\end{eqnarray}
\end{subequations}
with
\beq
  \lambda_f = e^{-2i\beta}\,  \frac{\mathcal{\bar M} (\bar B^0\to f)}{\mathcal{M} (B^0\to f)}
\label{lambdaf}
\eeq
and $\beta$ being the CKM matrix angle. Inserting the corresponding amplitudes $\mathcal{M} (\bar B^0\to f)$ from Eq.~\eqref{largeM} 
and $\mathcal{\bar M} (B^0\to f)$ from Eq.~\eqref{largeMbar} into Eqs.~\eqref{Sf} and \eqref{Af}, one obtains after integration over 
$\cos \theta$ the integrated $\mathcal{S}$ asymmetry 
\begin{eqnarray}
  \lefteqn{\hspace*{-3mm}\mathcal{S} (m_{\pi\pi}) =  (1-\mathcal{A})\, \times} \nonumber \\ 
  & \times & 2\, \mathrm{Im} \left \{ e^{-2i\beta}\left [ \,\bar a_S a_S^*+
      \frac{1}{3} |\mathbf{p}_\pi|^2  |\mathbf{p}_K|^2\, \bar a_P  a_P^* \right ] \right \}
\label{Sfint}
\end{eqnarray}
and the $\mathcal{A}$ asymmetry integrated over $\cos \theta$ is
\beq
  \mathcal{A} (m_{\pi\pi}) =\frac{|\bar a_S|^2-|a_ S|^2 + \frac{1}{3} |\mathbf{p}_\pi|^2  |\mathbf{p}_K|^2\left (|\bar a_P|^2 -|a_P|^2 \right )}
   {|\bar a_S|^2+|a_S|^2 + \frac{1}{3} |\mathbf{p}_\pi|^2  |\mathbf{p}_K|^2\left (|\bar a_P|^2 + |a_P|^2 \right )}.
\label{Afint}
\eeq

\section{Fitting procedure\label{sec4}}

In our fitting procedure we include the experimental data of Belle~\cite{Garmash:2005rv,Abe2005,Abe0509047,Abe0507037} and BaBar 
\cite{AubertPRD72,AubertPRD73,Aubert0408095,roKBaBar} Collaborations. They include altogether 18 values of branching fractions, direct 
$CP$ violating asymmetries for the charged  $B$ decays and the time dependent $CP$ asymmetry parameters $\mathcal{S}$ and $\mathcal{A}$ 
determined for the $B^0 \to \rho(770) K^0_S$ and $B^0 \to f_0(980) K^0_S$ decays. These data constitute the first part of the observable ensemble 
we fit. If one denotes by  $Z^\mathrm{exp}_i$ and $Z^\mathrm{th}_i$ the experimental and theoretical value of the above mentioned observables 
and by $\Delta Z^\mathrm{exp}_i$ their empirical errors, then the corresponding first part of the $\chi^2$ function reads: 
\beq
\chi^2_\mathrm{I}  = W_\mathrm{I}  \sum_{i=1}^{18}\left[\frac{Z^\mathrm{exp}_i - 
Z^\mathrm{th}_i}{\Delta Z^\mathrm{exp}_i}\right]^2.
\label{chi2def}
\eeq
Here, $W_\mathrm{I}=10$ is a weight factor chosen so as to make a good balance between this and the second component $\chi^2_\mathrm{II}$ of the 
total $\chi^2$ function. The $\chi^2_\mathrm{II}$ function is defined similarly to Eq.~\eqref{chi2def} but for the observables
 $Z_i$ being the 
$\pi\pi$ effective-mass and the helicity-angle distributions. We set $W_\mathrm{II}=1$. The experimental distributions have been background 
subtracted according to the figures of the Belle and BaBar papers cited above. For completeness, we give a list of figures showing the data we
used in the evaluation of $\chi^2_\mathrm{II}$: figures 5b, 6c, 6d and 7a till 7f from Ref.~\cite{Abe2005}, figures 5c, 6c, 6d from Ref.~\cite{Abe0509047}, 
figure 3b from Ref.~\cite{AubertPRD73}, and figure 3 from Ref.~\cite{AubertPRD72}. The experimental data are not absolutely normalized, therefore we 
have adopted an adequate method, described below, to compare the theoretical differential distributions of Eqs.~\eqref{mpipidecay.rate} and 
\eqref{dGamdcos} with the experimental $m_{\pi\pi}$ and $\cos \theta$ distributions.

\begin{table}[t]
\caption{Charming penguin parameters in the $S$-wave ($S_u$ and $S_t$) and in the $P$-wave ($P_u$ and $P_t$).}
\begin{tabular}{cc}
\hline
Parameter & Value   \\
\hline
$|S_u|$     & $0.15 \pm  0.10$ \\
Arg $S_u$   & $1.90 \pm 0.71$ \\
$|S_t|$     & $0.020 \pm 0.002$ \\
Arg $S_t$   & $-0.26 \pm 0.21$ \\
$|P_u|$     & $1.09 \pm 0.21$ \\
Arg $P_u$   & $-0.98 \pm 0.12$ \\
$|P_t|$     & $0.065 \pm 0.002$ \\
Arg $P_t$   & $-1.56 \pm 0.08$ \\
\hline
\end{tabular}
\label{tab:Par:results}
\end{table}

\begin{table*}[t!]
\caption{Average branching fractions $\mathcal{B}$ in units of $10^{-6}$, asymmetries $\mathcal{A}_{CP}$, and the asymmetry parameters 
$\mathcal{S}$ and $\mathcal{A}$ of our model compared to the values of Belle and BaBar Collaborations. The experimental branching 
ratios and their errors have been multiplied by 0.71 for $B \to \rho K $
channels and by 0.69 for $B \to f_0 K $ channels (for explanation 
see Section~\ref{sec4}).}
\begin{tabular*}{\textwidth}{p{60pt}p{70pt}p{80pt}p{80pt}p{60pt}p{40pt}p{60pt}c}
\hline
Observable & Channel & $\pi^+\pi^-$ mass & Our model & \multicolumn{2}{c}{Belle} &  \multicolumn{2}{c}{BaBar}  \\
 & & range (GeV) &   &   & Ref. &   & Ref.\\
\hline
 $\mathcal{B}$       & $\rho^0 K^\pm$ & $(0.66, 0.90)$ &$2.91  \pm 0.10$ &$2.76 \pm 0.45$  & \cite{Garmash:2005rv} &$3.60 \pm 0.71$ & \cite{AubertPRD72}\\ 
 $\mathcal{A}_{CP}$  & $\rho^0 K^\pm$ & $(0.66, 0.90)$ &$0.32  \pm 0.03$ &$0.30 \pm 0.14$  & \cite{Garmash:2005rv} &$0.32 \pm 0.16$ & \cite{AubertPRD72}\\ 
 $\mathcal{B}$       & $f_0 K^\pm$    & $(0.90, 1.06)$ &$6.93  \pm 0.16$ &$6.06 \pm 1.08$  & \cite{Garmash:2005rv} &$6.53 \pm 0.85$ & \cite{AubertPRD72}\\
 $\mathcal{A}_{CP}$  & $f_0 K^\pm$    & $(0.90, 1.06)$ &$0.02  \pm 0.02$ &\hspace{-0.25cm}$-0.08 \pm 0.08$  & \cite{Garmash:2005rv} &$0.09 \pm 0.12$ & \cite{AubertPRD72}\\
 $\mathcal{B}$       & $\rho^0 K^0$   & $(0.66, 0.90)$ &$3.49  \pm 0.20$ &$4.35 \pm 1.05$  & \cite{Abe0509047}     &$3.62 \pm 1.14$ & \cite{AubertPRD73}\\  
 $\mathcal{B}$       & $f_0 K^0$      & $(0.90, 1.06)$ &$3.86  \pm 0.14$ &$5.24 \pm 1.28$  & \cite{Abe0509047}     &$3.79 \pm 0.62$ & \cite{AubertPRD73}\\ 
 $\mathcal{S}$   & $\pi^+\pi^- K^0_S$ & $(0.66, 0.90)$ &$0.11  \pm 0.11$ &  $-$            &                       &$0.17 \pm 0.58$ & [26a]\\ 
 $\mathcal{A}$   & $\pi^+\pi^- K^0_S$ & $(0.66, 0.90)$ &$0.01  \pm 0.06$ &  $-$            &                       &\hspace{-0.25cm}$-0.64\pm 0.48$ & [26a]\\
 $\mathcal{S}$   & $\pi^+\pi^- K^0_S$ & $(0.89, 1.088)$&\hspace{-0.25cm}$-0.67 \pm 0.05$ &\hspace{-0.25cm}$-0.47\pm 0.37$ & [24a]     &       &   \\ 
 $\mathcal{S}$   & $\pi^+\pi^- K^0_S$ & $(0.86, 1.10)$ &\hspace{-0.25cm}$-0.63 \pm 0.05$ &                 &                       &\hspace{-0.25cm}$-0.95 \pm 0.29$ & \cite{Aubert0408095} \\ 
 $\mathcal{A}$   & $\pi^+\pi^- K^0_S$ & $(0.89, 1.088)$&\hspace{-0.25cm}$-0.11 \pm 0.07$ &\hspace{-0.25cm}$-0.23 \pm 0.26$ & [24a]     &      &   \\
 $\mathcal{A}$   & $\pi^+\pi^- K^0_S$ & $(0.86, 1.10)$ &\hspace{-0.25cm}$-0.11 \pm 0.07$ &                 &                       &\hspace{-0.1cm}$~0.24 \pm 0.34$ & \cite{Aubert0408095}\\

\hline
\end{tabular*}
\label{tab:Obs:results}
\end{table*}
 
For both differential distributions in $x=m_{\pi\pi}$ and  $x=\cos\theta$, we define the theoretical number of events in a bin $x_i$ by the integral
\beq
   Y_\mathrm{th} (x_i) =  \mathcal{N}\, \Gamma_B^{-1}  \int_{x_{i1}}^{x_{i2}} \frac{d\Gamma(x)}{dx} dx.
\eeq
The differential model distribution $d\Gamma (x)/dx$ is just that given by either Eq.~\eqref{mpipidecay.rate} or Eq.~\eqref{dGamdcos} and 
the integration range of the kinematic variables is the bin width $[ x_{i1}, x_{i2}]$. Then the normalization coefficient $\mathcal{N}$ 
of a given experimental distribution $Y_\mathrm{exp}$ is defined as a sum over all chosen bins:
\beq
 \mathcal{N} = \Gamma_B\, \frac{\sum_{i=1}^n Y_\mathrm{exp}
 (x_i)}{\int_{X_1}^{X_2}\!\dfrac{d\Gamma(x)}{dx} dx}.
\eeq
 The sum of experimental events $Y_\mathrm{exp} (x_i)$
over $x_i$ and the integration of $d\Gamma(x)/dx$ over $x$ is done in the analyzed range $[X_1, X_2]$ of interest, which must be
within the kinematically allowed range for the variables $m_{\pi\pi}$ and $\cos \theta$. Since we concentrate ourselves on the $\pi\pi$ 
effective mass range where the $\rho(770)^0$ and $f_0(980)$ resonances contribute the most, we take in our fits $X_1= 0.60$~GeV and 
$X_2 =1.06$~GeV.  

The experimental branching fractions for charged and neutral $B \to \rho(770)^0 K$ and $B \to f_0(980) K$ decays are obtained with the 
isobar model, in which the $B$-decay amplitude is a sum over the contributions of different two-body resonances. These are usually 
parameterized in terms of Breit-Wigner functions with the exception of the $f_0(980)$, for which the Flatt\'e formula is used. 
After fitting the intensities of the resonances, a particular branching fraction for a given resonance is calculated as an integral 
over the fully available phase space on the Dalitz plot. In order to compare these experimental branching fractions with those for 
limited effective mass ranges, we have calculated appropriate reduction coefficients as the ratio of the integral over  the limited 
phase space to that over the fully allowed one using the phenomenological amplitudes of the experimental analyzes. They are equal to 
0.71 in the case of the $\rho(770)^0$ and a $m_{\pi\pi}$ range $0.66-0.90$~GeV
and to 0.69 for the $f_0(980)$ which dominates in the 
$0.90-1.06$~GeV range. In fitting our model to the data, these coefficients reduce the branching fractions and their errors by the same factor. 

Experimental cuts on the effective pion-kaon mass influence the $\pi\pi$
effective-mass and the helicity angle distributions. A typical veto cut is experimentally used 
around the $K\pi$ effective mass, close to the $D$-meson mass. In our calculations, we have applied the cuts specified by the Belle and BaBar 
Collaborations in their analyzes of the $B$ decays into $\pi\pi K$.   

Our model has four complex parameters: $P_u$ and $P_t$ in the $P$-wave amplitude and $S_u$ and $S_t$ in the $S$-wave amplitude. 
With these parameters we shall describe more than two hundred data values.

\section{Results\label{sec5}}

We have performed a global fit to the available data measured by BaBar and Belle. These consist of 204 data points describing the $\pi\pi$ mass 
and angular distributions and of 18 observables enumerated in Section~\ref{sec4}. The $\chi^2$ values are the following: 
$\chi^2_\mathrm{I}/W_\mathrm{I} = 9.8$ and $\chi^2_\mathrm{II} = 336.3$. The values of the eight real parameters are listed in Table~\ref{tab:Par:results} 
along with the parabolic errors from the MINUIT minimization procedure~\cite{James:1975dr}.
\begin{figure*}[t!]
\includegraphics*[width=0.42\textwidth]{fig2a.eps} \hspace*{1.5cm}
\includegraphics*[width=0.42\textwidth]{fig2b.eps}
\caption{The $\pi^+\pi^-$ effective-mass distributions  {\bf (a)}  in $B^- \to \pi^+\pi^- K^-$ and {\bf (b)} in $B^+ \to \pi^+\pi^- K^+$ decays.
         The data are taken from the Belle Collaboration~\cite{Abe2005}. The solid lines represent the results of our model.} 
\label{fig2}  \bigskip
\includegraphics*[width=0.42\textwidth]{fig3a.eps}  \hspace*{1.5cm}
\includegraphics*[width=0.42\textwidth]{fig3b.eps}
\caption{Helicity-angle distributions in $B^\pm \to \pi^+\pi^- K^\pm$ {\bf (a)} in the $\rho(770)$ mass region (0.6 GeV $< m_{\pi\pi}<0.9$~GeV)
and {\bf (b)} in the $f_0(980)$ region (0.9~GeV $< m_{\pi\pi} < 1.06$~GeV). The data points are from Ref.~\cite{Abe2005} and the solid lines denote 
our model. The $S$-wave contribution is plotted with dashed lines, the  $P$-wave with dotted lines and the interference term with dot-dashed lines.}
\label{fig3}
\end{figure*}

\begin{figure}[t!]
\includegraphics*[width=0.42\textwidth]{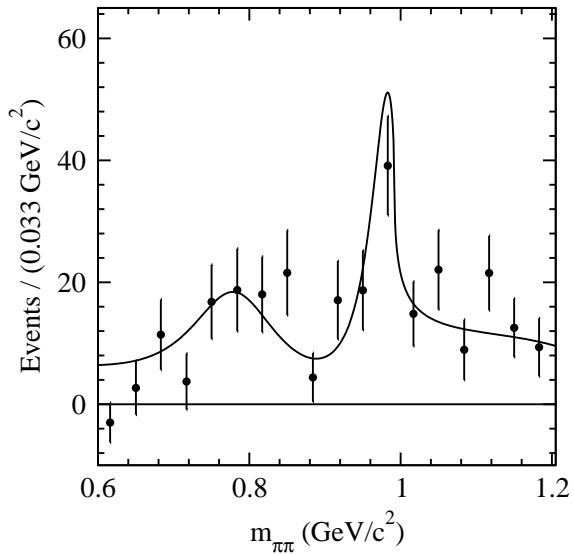}
\caption{The $\pi^+\pi^-$ effective-mass distributions in $B^0 \to \pi^+\pi^- K^0_S$ decays. Data points are taken from the BaBar Collaboration 
\cite{AubertPRD73} while the continuous line results from our model.}
\label{fig4}
\end{figure}

In Table~\ref{tab:Obs:results}, the values of branching fractions, direct and time-dependent $CP$ asymmetries are presented along with the 
corresponding experimental ones. The calculated observables for a given channel are integrated over the $m_{\pi\pi}$ range indicated in this table. 
Both the BaBar and Belle values are in agreement within their error bars and are well reproduced in our model. The theoretical errors stem from the 
parameter errors given in Table~\ref{tab:Par:results}. 

In Subsections~\ref{subA} and \ref{subB}, we present the $\pi\pi$ mass and helicity-angle distributions for charged and neutral $B\to\pi^+\pi^-K$ 
decays, respectively. Interference effects are studied in Subsection~\ref{subsec5c} and some discussion on the results is given in Subsection~\ref{subD}.

\begin{figure*}[t!]
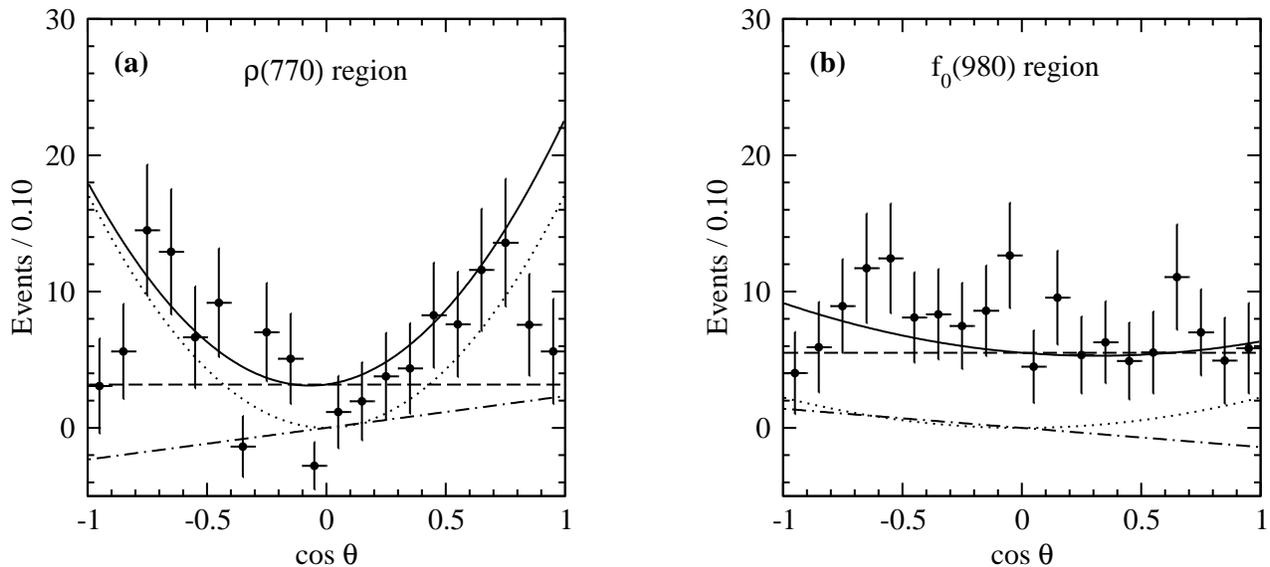

\includegraphics*[width=0.42\textwidth]{fig5a.eps} \hspace*{1.5cm}
\includegraphics*[width=0.42\textwidth]{fig5b.eps}
\caption{As in Fig.~\ref{fig2} but for $B^0 \to \pi^+\pi^- K^0_S$ decays and for data from Belle~\cite{Abe0509047}.}
\label{fig5}
\end{figure*}

\subsection{The $\bm{\pi\pi}$ mass and helicity-angle distributions in $\bm{B^\pm\to \pi^+\pi^- K^\pm}$ decays \label{subA}}

In Fig.~2 the $\pi\pi$ effective-mass distributions of our model are compared to the Belle data~\cite{Abe2005}. In the fit to these distributions,
the background corrected data have been used. One sees from Figs.~2a and 2b that our model describes rather well the $\pi^+\pi^-$ spectra 
measured by the Belle Collaboration separately for the $B^+$ and $B^-$ decays. Our theoretical curve depicts, as in previous work~\cite{fkll}, 
a prominent maximum near 1~GeV, to which now adds a less pronounced peak at about $770-780$~MeV. Remarkably, one observes a clearly 
visible asymmetry in number of events between the $B^-\to K^-\pi^+\pi^-$ and $B^+\to K^+\pi^+\pi^-$ decays for the $\rho(770)^0$ and $f_0(980)$ 
regions. At lower $m_{\pi\pi}$ masses about 500~MeV, our model also produces a broad maximum which we attribute to the $\sigma$ or $f_0(600)$. 
By the same token, we note that in fits to their data, the Belle Collaboration has not included the $\sigma$ resonance which is thus buried in 
the non-resonant background. Let us remark that this part of the $m_{\pi\pi}$ spectrum has not been fitted in our calculations. As explained 
in the previous section~\ref{sec4}, we have intentionally chosen the lower mass limit to be equal to 0.6~GeV.  

In Fig.~3, we compare the results of our model for the $\cos \theta$ distribution in the vicinity of the $\rho(770)^0$ (Fig.~3a) and of the 
 $f_0(980)$ (Fig.~3b) with the Belle data~\cite{Abe2005}. The experimental $\cos\theta$ distribution is fairly well reproduced given the fluctuation 
in events of the data points. The behavior of the angular distribution such as seen near the $f_0(980)$ mass is typical for a substantial interference 
pattern between the $S$- and $P$-wave. The former gives a constant contribution while the latter is a symmetric $\cos^2\theta$ distribution 
observed in the $\rho(770)^0$ mass range (see Eq.~\eqref{dGamdcos}). The $S-P$ interference term, which is proportional to $\cos \theta$, is clearly 
seen in Fig.~3b. This demonstrates the presence of the $P$-wave contribution in the $f_0(980)$ region. 

\begin{figure*}[t!]
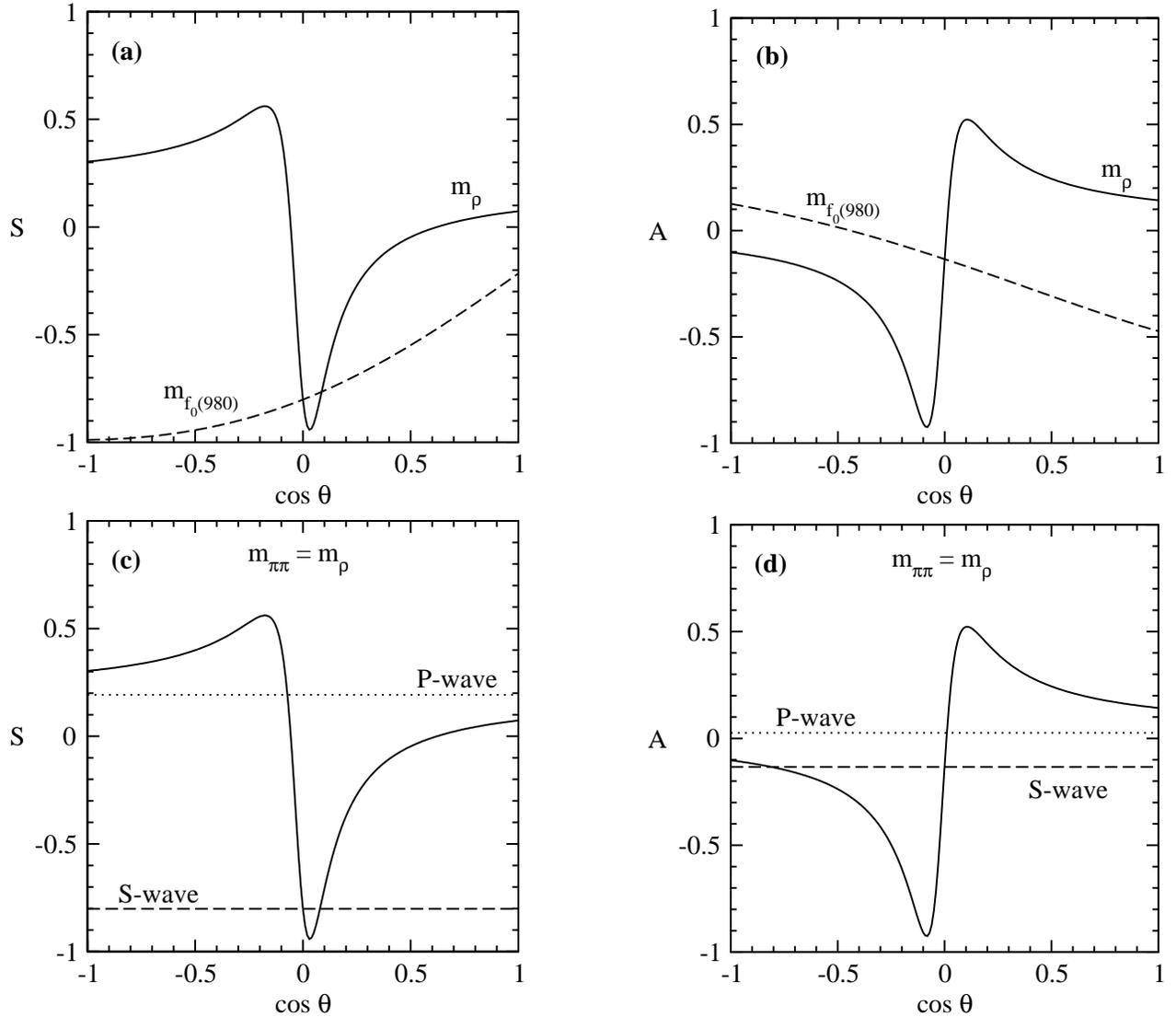

\includegraphics*[width=0.42\textwidth]{fig6a.eps}  \hspace*{1.5cm}
\includegraphics*[width=0.42\textwidth]{fig6b.eps} \\
\includegraphics*[width=0.42\textwidth]{fig6c.eps}  \hspace*{1.5cm}
\includegraphics*[width=0.42\textwidth]{fig6d.eps}
\caption{Helicity-angle dependence of the $CP$ violating asymmetries $\mathcal{S}$ and $\mathcal{A}$ calculated for $B^0 \to \pi^+\pi^- K^0_S$ 
decays at $m_{\pi\pi}$ equal to $m_\rho = 775.8$~MeV (solid lines in {\bf (a)}--{\bf (d)}) and to $m_{f_0(980)} = 980$ MeV (dashed lines in {\bf (a)} 
and {\bf (b)}). Separate $S$-wave (dashed lines) and $P$-wave results (dotted lines) are shown in {\bf (c)} and {\bf (d)} for $m_{\pi\pi} = m_\rho$.}
\label{fig6}
\end{figure*}

\begin{figure*}[t!]
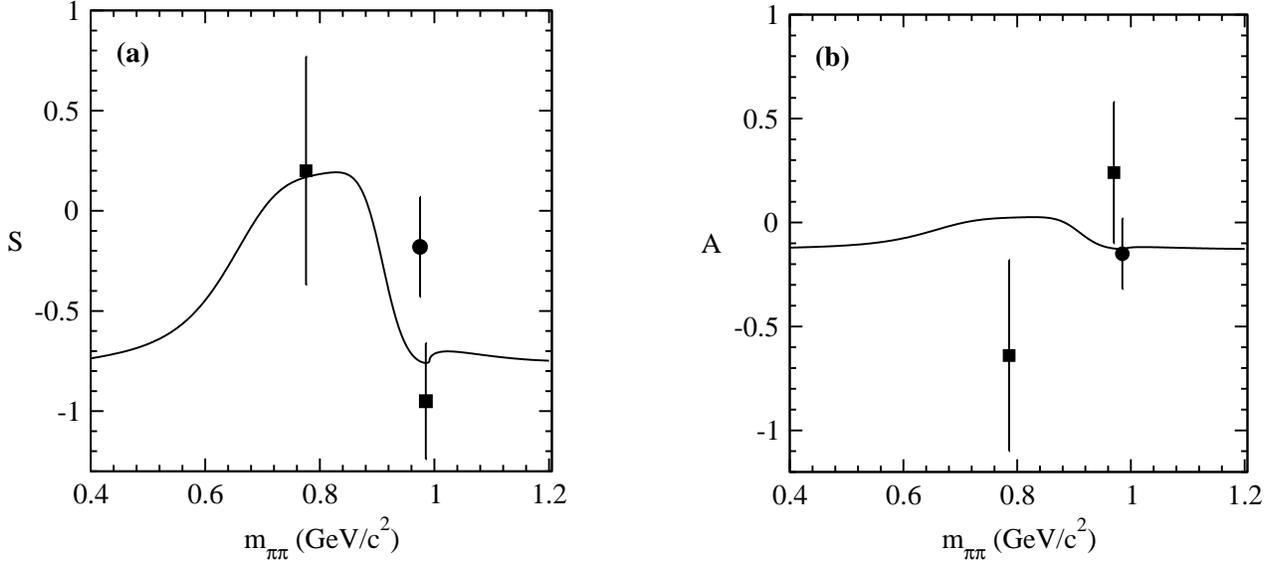

\includegraphics*[width=0.42\textwidth]{fig7a-new.eps}  \hspace*{1.5cm}
\includegraphics*[width=0.42\textwidth]{fig7b-new.eps}
\caption{Effective-mass dependence of the $CP$ violating asymmetries, integrated over $\cos \theta$, $\mathcal{S}$ in {\bf (a)} and $\mathcal{A}$ 
in {\bf (b)} obtained with our model (solid lines). The Belle Collaboration results [24b] for the $f_0(980)$ resonance are plotted as 
circles and the BaBar data for the $f_0(980)$~\cite{Aubert0408095} and $\rho(770)^0$~[26b] as squares.}
\label{fig7}
\end{figure*}

\subsection{The $\bm{\pi\pi}$ mass and helicity-angle distributions in $\bm{B^0\to \pi^+\pi^- K^0}$ decays \label{subB}}

The $m_{\pi\pi}$ effective mass distribution for neutral $B$ decays is plotted in Fig.~4 along with the results obtained by BaBar~\cite{AubertPRD73}. 
Within the error bars our model also describes this $m_{\pi\pi}$ distribution well, which bears the same features concerning the $\rho(770)^0$ 
and $f_0(980)$ as the charged $B$ decay $m_{\pi\pi}$ spectrum. In Fig.~5a and 5b the $\cos \theta$ distributions in the proximity of the $\rho(770)^0$ 
and $f_0(980)$ are compared with the experimental data from Belle~\cite{Abe0509047}. While the $\rho(770)^0$ range is dominated by 
the $P$-wave contribution, in the $f_0(980)$ range one observes once again the interference pattern. The sign of the interference term is, however, 
opposite to that apparent in Fig.~2b. It is well explained by our model in which the $P$-wave and $S$-wave amplitudes for both charged and neutral 
$B$ decays are closely related (see Eqs.~\eqref{simplification.beneke} and \eqref{simplification.beneke0} where the signs of the charming 
penguin terms are reversed).

\subsection{Interference effects\label{subsec5c}}

In this subsection we study the interplay between the $S$- and $P$-waves of the decay amplitudes. Fig.~\ref{fig6} shows the angular dependence 
of the two $CP$ violating asymmetries $\mathcal{S}$  and $\mathcal{A}$ calculated at the effective masses $m_{\pi\pi}=775.8$~MeV and 
$m_{\pi\pi}=980$~MeV using Eqs.~\eqref{Sf} and \eqref{Af} and \eqref{lambdaf}. These numbers are chosen to demonstrate the behavior of the asymmetries 
about the $\rho(770)^0$ and $f_0(980)$ resonances. As seen in Fig.~6a and 6b, the functions $\mathcal{S}$ and $\mathcal{A}$ are not constant. 
This is a clear demonstration of the interference between $S$- and $P$-waves. The sharp changes in behavior of the asymmetry $\mathcal{S}$ 
at the $\rho(770)^0$ mass are shown in Fig.~6c, where the straight dashed and dotted lines are the $\mathcal{S}$ values for separate $S$- 
and $P$-wave contributions, respectively. At $\cos \theta =0$, the $P$-wave amplitude vanishes, therefore the strongest variations of the 
asymmetry are observed near $\theta = \pi /2$. A similar picture for the asymmetry $\mathcal{A}$ is shown in Fig.~6d. 

The asymmetries $\mathcal{S}$  and $\mathcal{A}$ at $m_{\pi\pi}= 980$~MeV behave smoothly near $\cos \theta =0$ since there the 
$S$-wave dominates. A decrease of $\mathcal{A}$ as well as an increase of $\mathcal{S}$ as functions of $\cos \theta$ are due to the 
$P$-wave amplitude component, which is still non-negligible in the $f_0(980)$ mass range.  

Recall that the helicity angle dependence of the asymmetries can be transformed
into the functional dependence on the $\pi^- K$ effective mass
\beq
   m_{\pi^- K}= \sqrt{m_{\pi}^2 +M_{K}^2+m_{\pi\pi}E_K(m_{\pi\pi}) + 2\, |\mathbf{p}_{\pi}||\mathbf{p}_K| \cos \theta},   
\eeq
where $m_\pi$ is the pion mass. Thus, the above interference effects can also be experimentally studied by examination of particular regions 
of the Dalitz plot. 

In Fig.~\ref{fig7}, the effective $\pi\pi$ mass dependence of the asymmetries integrated over $\cos \theta$ are plotted (see Eqs.~\eqref{Sfint} 
and \eqref{Afint}). The most pronounced effect is an intermediate change of sign of $\mathcal{S}$ near the $\rho(770)^0$ resonance. 
We stress that if the $S$- or $P$-amplitudes were dominated by only one weak amplitude proportional to $\lambda_t$ then the asymmetry 
$\mathcal{S}$ shown in Fig.~7a would suddenly jump from $-\sin 2\beta \approx - 0.7$ at the $\pi\pi$ threshold to $+ \sin 2\beta \approx +0.7$ 
near the $\rho(770)^0$ mass and abruptly drop down to $-\sin 2\beta \approx - 0.7$ at the $f_0(980)$ mass, as can be inferred from Eq.~\eqref{Sfint}. 
A smooth behavior of $\mathcal{S}$ is explained by the mass dependence of the $S$-wave scalar form factors and by the finite width of the
$\rho(770)^0$.  As seen in Fig.~7a, our model is in agreement with the data. Remarkably, there is a particularly large departure from the 
$+ \sin 2\beta$ value near the $\rho(770)^0$ mass. This can be explained by a substantial value of the weak part of the amplitude in 
Eqs.~\eqref{def:C} and \eqref{simplification.beneke0} proportional to $\lambda _u$. In this context, bear in mind that the $P$-wave penguin 
parameter $|P_u|$ is much larger than the $S$-wave parameter $|S_u|$ (see Table~I). 

The functional dependence of the asymmetry $\mathcal{A}$ on $m_{\pi\pi}$ is depicted in Fig.~7b. It is not as strong as for the $\mathcal{S}$ 
asymmetry, but certainly $\mathcal{A}$ is not equal to zero. In the $f_0(980)$ range, $\mathcal{A}$ is close to $-0.1$. Hence, it is comparable 
to the values of $\mathcal{A}$ found by Belle and Babar in $B^0 \to K^+ \pi^-$ decays~\cite{baba04pik,bellepik}. In the former case, however, 
the experimental errors are yet too large to claim the non-zero value of $\mathcal{A}$ (see Table II). 

We conclude this section by remarking that in penguin dominated decays, such as $B\to \rho K, \omega K ... $, tree diagrams are either CKM 
suppressed or CKM and color suppressed. Still, we observe that for $\rho(770)^0 K$ final states an accidental cancellation occurs in the 
$\lambda_u$ terms of Eqs.~\eqref{eq:U} and \eqref{eq:U0} due to the combinations $a_4^c - r a_6^c$ and $a_4^u-ra_6^u$, 
where $r\simeq 1, a^{u,c}_6\approx a^{u,c}_4$. A similar cancellation takes equally place in the $B\to f_0(980)K$ channel~\cite{fkll}. 
Therefore, the `tree pollution' is considerable and one may, at this point already, expect deviations from the value $\mathcal{S} \approx \sin 2\beta$ 
associated with the pure tree diagram for the $B^0\to J/\psi K_S$ or with the penguin diagram for the $B^0\to \phi K^0$ decays. The tree diagram 
contribution to the $\rho(770)^0K$ state, however, is not sufficient to explain the important departure of $\mathcal{S}$ from the above value. 
It is the aforementioned long-distance contribution $|P_u|$ to the $\lambda_u$ term combined with interference effects that determine the 
functional behavior of the $\mathcal{S}$ asymmetry parameter.

\subsection{Discussion of the results \label{subD}}

We stress the importance of charming penguin terms. Without them it is not possible to obtain a good agreement of the theoretical branching ratios 
with experimental data. If all the penguin parameters are set equal to zero then the model branching ratio is underestimated by a factor of about 
5 for the $B^0 \to f_0(980) K^0$ decay and by 3.5 for the $B^0 \to \rho(770)^0 K^0$ decay. For the charged $B$ decays the theoretical branching ratios 
are too small by factors of about 2.4 and 7 for the $f_0(980) K$ and $\rho(770)^0 K$ channels, respectively. 

It has, however, been found in Ref.~\cite{Cheng:2005nb} that the branching ratios for $B\to f_0(980)K$ decays can be reproduced in the QCD factorization 
approach provided that one uses a large scalar decay constant $f_{f_0}^s$ of 370~MeV. The work of Ref.~\cite{Cheng:2005nb} is based on a two-body 
$B$-decay amplitude. Our approach considers a three-body $B$-decay amplitude by taking into account the $f_0(980)$ disintegration into two pions or 
two kaons and also by introducing final-state interactions between them. A relation between the two- and three-body $B$-decay amplitudes, elucidated 
in the Appendix~\ref{appendix}, leads in our case to a value of 94~MeV for $f_{f_0}^s$ as seen in Eq.~(A14). The difference with the value of 
Ref.~\cite{Cheng:2005nb} by a factor of about 4 explains partly why we underestimate the $B\to f_0(980)K$ branching ratio if long-distance charming 
penguin contributions are absent.

We have studied the sensitivity of our results on the scalar form factors $\Gamma_{i=1,2}^{n,s}(m_{\pi\pi})$ which depend on the not very well known 
low-energy constants of the chiral perturbation theory. The possible variation of their values (see Refs.~\cite{Meissner:2000bc} and \cite{laehde06}) 
can lead up to a multiplicative factor of about 1.25 for $\Gamma_{i=1,2}^{n,s}(m_{\pi\pi})$ in the $f_0(980)$ range. Note however, that the sensitivity 
to such changes is limited due to the constant value of the product $\chi\vert \Gamma_1^n(m_{f_0})\vert$ given by Eq.~(A15). If, for example, 
$\Gamma_1^n(m_{f_0})$ is multiplied by a factor of 1.25, the $\chi$ value is to be divided by 1.25 and 
consequently the value of $f_{f_0}^s$ is also multiplied by 1.25 (see Eq. (A14)). 
Thus, an increase of the scalar form factor  $\Gamma_1^s(m_{f_0})$ will only enhance the contribution of the $Q(m_{\pi\pi})V$ term in Eq.~(A12).

Our $B\to(\pi\pi)_SK$ amplitude depends also on the $F_0^{B\to(\pi\pi)_S}(M_K^2)$ transition form factor which is a model dependent quantity.
Decreasing its value from 0.46 (see Subsection~\ref{sec2c}) to 0.25 (value quoted in Ref.~\cite{Cheng:2005nb}) will lead to a readjustment of the charming 
parameters $S_u$ and $S_t$ while fitting the experimental data. The quality of the fit is good and similar to that presented here.
We have also obtained a good fit using the values of the low energy constants determined in Ref.~\cite{laehde06} (the values of the first line of 
Table I therein) for the scalar form factors together with $F_0^{B\to(\pi\pi)_S}(M_K^2)=0.25$.  Our conclusions are not changed 
by these possible modifications of our input.

\section{Conclusions and Outlook \label{sec6}}

In our studies of $B \to \pi^+\pi^- K$ decays, we have extended the work of Ref.~\cite{fkll} by adding the $P$-wave contribution to that of the 
$S$-wave in the two-pion final-state interactions. The total amplitude contains the resonant channels $B \to \rho(770)^0 K$ and $B \to f_0(980) K$.
This work goes beyond the usual approach of calculating two-body branching ratios of the $B$ decays, as we have taken into account 
parts of the final state interactions between particles in three-body decay channels. In this paper, we have analyzed the interactions between 
pairs of pions in the Dalitz plot  section with the $\pi\pi$ effective mass range from threshold to about 1.2~GeV. In the $S$-wave the rescattering or 
the transition amplitudes between two pions or two kaons are described by the unitary coupled-channel model of Ref.~\cite{Kaminski:1997gc}. 

There are two components in the weak transition amplitudes. The first term is derived within the factorization approximation with some QCD corrections  
without hard-scattering and annihilation terms. The second contribution, called charming penguins, is a long-distance amplitude originating from 
penguin-type diagrams with $c$-quark or $u$-quark loops.  Four  complex charming penguin parameters, common for $B^+, B^-, B^0$ and $\bar B^0$ 
decays, have been introduced and fitted to numerous experimental data, including the $\pi\pi$ effective mass and helicity angle distributions, branching 
fractions, direct asymmetry $\mathcal{A}_{CP}$ and time dependent $CP$ violating asymmetry parameters $\mathcal{S}$ and $\mathcal{A}$. 
Our theoretical model reproduces well the experimental results. Without the charming penguin amplitudes, the model branching fractions of 
$B \to f_0(980) K$ and $B \to \rho(770)^0 K$ decays are several times smaller than the experimental values. In the $m_{\pi\pi}$ range below 1.2~GeV, 
the two main resonances $\rho(770)^0$ and $f_0(980)$ give rise to important interference effects, best visible in the helicity-angle distributions. 
Thus, in  Figs.~\ref{fig3} and \ref{fig5}, we notice sizable interference terms of opposite signs comparing the $B^\pm \to \pi^+\pi^- K^\pm$ decays with 
the $B^0 \to \pi^+\pi^- K^0$ ones. 

In our model,  the  direct $CP$ asymmetry in $B^\pm \to \rho^0 K^\pm$ decays is in agreement with the large experimental value measured by the Belle and 
BaBar groups. The parameter $\mathcal{S}$ of the time-dependent $CP$ asymmetry for $B^0 \to \rho^0 K^0_S$ decays is significantly smaller than 
the value $\sin 2 \beta$ expected in case of a full dominance of the weak decay
amplitude proportional to $\lambda_t$. This is related to a large value of the $u$-penguin parameter 
$P_u$ in the $P$-wave (see Table~\ref{tab:Par:results}). The numbers $S_u$, $S_t$, $P_u$ and $P_t$ should be treated as phenomenological parameters 
which we attribute to the long-distance penguin contributions. In our opinion, however, they can  also contain contributions from other processes omitted by 
us like annihilation or hard-spectator interaction terms.

In this approach, all the resonances appear in a natural way as poles of the meson-meson amplitudes. No arbitrary phases nor any relative intensity 
parameters for the two discussed  resonances $\rho(770)^0$ and $f_0(980)$ are needed. This is in contrast to the usual phenomenological analyzes in 
terms of the isobar model applied to fit the Dalitz plot density distributions. Let us remark that the scalar resonance $f_0(600)$, as a pole of the 
$S$-wave amplitude, is also present in the $\pi\pi$ effective mass spectrum. However, it has not been included in the phenomenological models of the 
Belle and BaBar Collaborations~\cite{Garmash2005,AubertPRD72,Abe2005}.

So far, we have restricted our analysis to an effective $\pi\pi$ mass of about $1.2$~GeV. In further studies, one can extend not only this mass 
range to include further scalar and vector resonances but also treat the other variable $m_{\pi K}$ on the Dalitz plot and the accompanying vector 
resonances $K^*$. Given our findings for the $f_0(980)$ and $\rho(770)^0$ resonances, we expect many more interference effects at larger 
effective $\pi\pi$ masses and also for the $\pi K$ effective mass range.

\begin{acknowledgments}
One of us (L.~L.) is very grateful to Maria R\'o\.za\'nska for many useful conversations and remarks on $B$-meson decays. B.E. is thankful 
to Thomas Latham for providing access to the BaBar raw data on effective mass distributions including background corrections. 
B.E. and B.L. acknowledge pleasant and helpful discussions with Jos\'e Ocariz and also thank Olivier Leitner for quite useful comments. 

This work has been performed within the framework of the IN2P3-Polish Laboratories Convention (Project No. CSI-12). A visit of B.~E. in 
Krak\'ow has been partly financed within an agreement between the CNRS (France) and the Polish Academy of Sciences (Project No. 19481).   
B.~E. is supported by a Marie Curie International Reintegration Grant under 
the Contract No. 516228. 
\end{acknowledgments}

\appendix
\section{Amplitudes for the $\bm{B^-\to f_0(980)K^-}$ and $\bm{B^-\to(\pi^+\pi^-)_S K^-}$ decays \label{appendix}}

We give here the expressions for the two-body  $\langle f_0K^-\vert H\vert B^-\rangle$ and the three-body $\langle(\pi^+\pi^-)_SK^-\vert H\vert B^-\rangle$ 
decay amplitudes and show how they are related to each other. In the QCD factorization approach, adding charming penguin terms, the former amplitude 
can be written as
$$
\begin{array}{l}
$$
\label{eq:A1}
      \langle f_0K^-\vert H\vert B^-\rangle =\dfrac{G_F}{\sqrt{2}}
      \left[ P(m_{f_0})U+C^S(m_{f_0})\right.+\\
\hspace{3cm} +  \left.\bar Q(m_{f_0})V+
  C^S(M_K)\right] \ .
\end{array}
\eqno{(A1)}
$$
If electroweak penguin and annihilation terms are not included,
$$
\label{eq:A2}
P(m_{f_0})=f_K (M_B^2-m_{f_0}^2)F_0^{B\to f_0}(M_K^2)\ ,
\eqno{(A2)}
$$
$$
\label{eq:A3}
\bar Q(m_{f_0})\!=\!\frac{2\langle f_0\vert \bar ss\vert 0\rangle}{m_b\!-\!m_s}(M_B^2\!-\!M_K^2)F_0^{B\to K}(m_{f_0}^2)\ ,
\eqno{(A3)}
$$
$$
\label{eq:A4}
U\!\!=\!\lambda_u\!
\left[
a_1\!+\!a_4^{u}\!\!-\!a_4^c\!+\!\left(a_6^c\!-\!a_6^{u}\right)r
\right]
+\lambda_t\left(a_6^cr\!-\!a_4^c\right)\ ,
\eqno{(A4)}
$$
$$
\label{eq:A5}
V=\lambda_u\left(a_6^c-a_6^{u}\right)+\lambda_ta_6^c\ ,
\eqno{(A5)}
$$ 
where $m_{f_0}$ is the $f_0(980)$ mass and $F_0^{B\to f_0}(M_K^2)$  and $F_0^{B\to K}(m_{f_0}^2)$ are the $B\to f_0(980)$ and $B\to K$ 
transition form factors. The scalar decay constant $f_{f_0}^s$ can be defined by the matrix element
$$
\label{eq:A6}
\langle f_0\vert \bar ss\vert 0\rangle=m_{f_0}f_{f_0}^s\ .
\eqno{(A6)}
$$ 
The three-body decay amplitude reads
$$
\begin{array}{l}
\label{eq:A7}
      \langle(\pi^+\pi^-)_SK^-\vert H\vert B^-\rangle \\
     = \dfrac{G_F}{\sqrt{2}} \sqrt{\frac{2}{3}}\Big\{ 
      \left[ P(m_{\pi\pi})U+C^S(m_{\pi\pi})\right] \Gamma_{f_0\pi\pi}^n(m_{\pi\pi})+\\
 \hfill +
  \left[\bar Q(m_{\pi\pi})V+
  C^S(M_K)\right] \Gamma_{f_0\pi\pi}^s(m_{\pi\pi})\Big\},
\end{array}
\eqno{(A7)}
$$
where $\Gamma_{f_0\pi\pi}^n(m_{\pi\pi})$ and $\Gamma_{f_0\pi\pi}^s(m_{\pi\pi})$ are the non-strange and strange vertex functions, respectively.
They describe the decay of the $f_0(980)$ into two pions. The factor $\sqrt{2/3}$ is the Clebsch-Gordan coefficient which relates the isospin-0 
$S$-wave $\vert(\pi\pi)_S\rangle$ to the $\vert(\pi^+\pi^-)_S\rangle$ one. The vertex function $\Gamma_{f_0\pi\pi}^s(m_{\pi\pi})$ is defined 
through the relation between the matrix elements
$$
\label{eq:A8}
\langle (\pi\pi)_S\vert \bar ss\vert 0\rangle=\Gamma_{f_0\pi\pi}^s(m_{\pi\pi})\langle f_0\vert \bar ss\vert 0\rangle\ .
\eqno{(A8)}
$$
Replacement of $\bar ss$ in Eq.~(A8) by $\bar nn=\frac{1}{\sqrt{2}}(\bar uu+\bar dd)$ yields the corresponding $\Gamma_{f_0\pi\pi}^n(m_{\pi\pi})$ 
vertex function. The matrix element $\langle (\pi\pi)_S\vert \bar ss\vert 0\rangle$ is related to the strange scalar form factor 
$\Gamma_1^s(m_{\pi\pi})$ by \cite{Meissner:2000bc}
$$
\label{eq:A9}
\langle (\pi\pi)_S \vert \bar ss\vert 0\rangle=\sqrt{2}B_0\Gamma_1^{s*}(m_{\pi\pi})\ ,
\eqno{(A9)}
$$
where $B_0=-\langle0\vert \bar qq\vert 0\rangle/f_\pi^2$ is proportional to the quark condensate, $f_\pi$ being the pion decay constant.
Let us note, by this occasion, that in Eq.~(A9) the charge conjugation appears since the form factor $\Gamma_1^s$ is defined as being 
proportional to the matrix element $\langle 0\vert \bar ss\vert  (\pi\pi)_S \rangle$. One must also correct Eq.~(11) of Ref.~\cite{fkll} in which
$\Gamma_i^{n,s}$ should be replaced by ${\Gamma_i^{n,s}}^*$. Using Eqs. (A6, A8, A9) one finds
$$
\label{eq:A10}
\Gamma_{f_0\pi\pi}^s(m_{\pi\pi})=\chi\ \Gamma_1^{s*}(m_{\pi\pi})\ ,
\eqno{(A10)}
$$
where the constant $\chi$ is defined by
$$
\label{eq:A11}
\chi=\frac{\sqrt{2}B_0}{m_{f_0}f_{f_0}^s}\ .
\eqno{(A11)}
$$
We assume $\Gamma_{f_0\pi\pi}^n(m_{\pi\pi})$ to be related to the non-strange scalar for factor $\Gamma_1^{n*}(m_{\pi\pi})$ by the same 
constant $\chi$ as in Eq.~(A10). Here, $\Gamma_1^{n*}(m_{\pi\pi})=\langle (\pi\pi)_S\vert \bar nn\vert 0\rangle/\sqrt{2}B_0$ similarly to Eq.~(A9).
Making the appropriate replacements for $\Gamma_{f_0\pi\pi}^s(m_{\pi\pi})$ and $\Gamma_{f_0\pi\pi}^n(m_{\pi\pi})$ in Eq.~(A7), we finally arrive at
$$
\begin{array}{l}
\label{eq:A12}
      \langle(\pi^+\pi^-)_SK^-\vert H\vert B^-\rangle \\
     =\dfrac{G_F}{\sqrt{2}}\sqrt{\frac{2}{3}}\Big\{ \chi
      \left[ P(m_{\pi\pi})U+C^S(m_{\pi\pi})\right] \Gamma_1^{n*}(m_{\pi\pi})+\\
 \hfill +
  \left[Q(m_{\pi\pi})V+\chi
  C^S(M_K)\right] \Gamma_1^{s*}(m_{\pi\pi})\Big\}\ ,
\end{array}
\eqno{(A12)}
$$
where the relation
$$
\label{eq:A13}
Q(m_{\pi\pi})=\bar Q(m_{\pi\pi})\chi
\eqno{(A13)}
$$
holds. Thus, Eq.~(A12) is identical to Eq.~(1) of Ref.~\cite{fkll}. From Eq.~(A11) it follows
$$
\label{eq:A14}
f_{f_0}^s=\frac{\sqrt{2}B_0}{m_{f_0}\chi}.
\eqno{(A14)}
$$
With $\chi=29.8$ GeV$^{-1},\ B_0=m_\pi^2/2\hat m$ and $\hat m=5$ MeV, $m_{f_0}=980$ MeV one obtains $f_{f_0}^s=94$ MeV. 
In our model $f_{f_0}^s$ is determined by the $f_0(980)$ properties since \cite{fkll}, 
$$
\chi=\frac{g_{f_0\pi\pi}}{m_{f_0}\Gamma_{tot}(f_0)}\ \frac{1}{\vert\Gamma_1^n(m_{f_0})\vert}\ ,
\eqno{(A15)}
$$
here $\Gamma_{tot}(f_0)$ is the total $f_0(980)$ width.
The constant $\chi$ enters in the description of the decay of the scalar resonances into two pions as seen from Eq.~(A10) and comments below Eq.~(A11).
Furthermore Eq.~(A15) shows that $\chi$ has the same role as the coupling constant $g_{\rho\pi\pi}$ of Eq.~(A14).

As can be seen from Eqs.~(A1) and (A6) the two-body part of QCD factorization amplitude without charming penguins is similar to the corresponding one of 
Ref.~\cite{Cheng:2005nb} when excluding electroweak penguin and annihilation contributions.
Note, however, that we disagree with the relative sign between the $P(m_{\pi\pi})U$ (Eqs. (A2) and (A4)) and $Q(m_{\pi\pi})V$ (Eqs. (A3) and (A5)) terms.
Our relative sign agrees with that of Ref.~\cite{Delepine}.

 \end{document}